\documentclass[11pt]{article}
\usepackage[utf8]{inputenc}
\usepackage{hyperref}
\usepackage{graphicx}% Include figure files
\usepackage{dcolumn}% Align table columns on decimal point
\usepackage{bm}% bold math
\usepackage{longtable}
\usepackage{amsmath}
\usepackage{multirow}
\usepackage{lineno,rotating}
\usepackage{geometry}
 \usepackage{epstopdf}

\RequirePackage{wasysym} \RequirePackage{setspace}
\pdfoutput=1
\begin{document}

\title{\bf Excited states of odd-mass nuclei with different deformation-dependent  mass coefficients}
\author{M. Chabab$^{1}$, A. El Batoul$^{1}$, I. El-ilali$^{1}$, A. Lahbas$^{2,1}$  and M. Oulne$^{1,*}$\\
\\ {\small $^{1}$ High Energy Physics and Astrophysics Laboratory, Faculty of Sciences Semlalia, }\\
{\small Cadi Ayyad University, P. O. B. 2390, Marrakesh 40000, Morocco} \\
{\small $^{2}$ESMaR, Department of Physics, Faculty of Sciences, Mohammed V University in Rabat, Morocco} \\
{\small $^{*}$ corresponding author : oulne@ucam.ac.ma} \\
}

%\\ \small $^{1}$ High Energy Physics and Astrophysics Laboratory, Department of %Physics,  Faculty of Sciences Semlalia, Cadi Ayyad University,
%P.O.B. 2390, Marrakesh 40000, Morocco. \\
%\small $^{2}$ ESMaR, Department of Physics, Faculty of Sciences, Mohammed V University in Rabat, Morocco.
%{\small $^{*}$ corresponding author : oulne@ucam.ac.ma}%

\maketitle
\abstract{
Experimental data indicate that the mass tensor of collective Bohr Hamiltonian cannot be considered as a constant but should be considered as a function of the collective coordinates. In this work our purpose is to investigate the properties of  low-lying collective states of the odd nuclei $^{173}$Yb and $^{163}$Dy by using  a new generalized version of the collective quadrupole Bohr Hamiltonian with deformation-dependent  mass coefficients.
The proposed new version of the Bohr Hamiltonian  is solved for Davidson potential in $\beta$ shape variable, while the $\gamma$ potential is taken to be equal to the harmonic oscillator.  The obtained results of the excitation energies and B(E2) reduced transition probabilities show an overall agreement with the experimental data. Moreover, we investigate the effect of the deformation dependent mass parameter on energy spectra and transition rates in both cases, namely: when the mass coefficients are different and when they are equal. Besides, we will show the positive effect of the present formalism on the moment of inertia.
%
     % end of PACS codes
} %end of abstract
\section{Introduction}
\label{intro}
 Quantum Phase Transitions \cite{Cejnar,Iachello2000,Casten2000} in atomic nuclei within  the Bohr-Mottelson Model (BMM) \cite{norman,2,Buganu} have attracted a considerable attention for describing  the  quadrupole collective excitations behavior in various deformed nuclei. In this context, the Bohr Hamiltonian involved in this model has a standard form of the kinetic energy term which contains one  mass coefficient for all modes of excitation, namely : the $\beta$- and the $\gamma$- vibrations and the rotational motion, where $\beta$  is the collective variable
corresponding to nuclear deformation while $\gamma$ denotes an angle measuring departure from axial symmetry \cite{3}. This approximation is argued in terms of small oscillations around the equilibrium value. However, several authors have elaborated  new approaches to generalize the usual form of the kinetic energy term of the Bohr Hamiltonian given in the intrinsic frame. One can cite two remarkable ways  : The first one is the approach followed by Jolos and von Brentano \cite{7,71,8}, in which they showed, for low lying collective states of well-deformed axially symmetric even-even nuclei, the necessity of the introduction of three different mass coefficients for each collective mode motion (ground state, $\beta$ or $\gamma$). These latter are determined by the experimental data on B(E2)'s and the excitations energies. This approach was applied in Ref. \cite{9} with Davidson potential, and has  shown a strong influence of different mass parameters, especially on the interband B(E2) rates. Also,  the high-spin states of spectra and intraband B(E2) are affected by these differences. However, it is  worth to notice here that the used energy formula in Ref. \cite{9} is inaccurate as we have already shown in \cite{14} for even-even nuclei and that we will show once again in the present work for odd-A nuclei.That is why our calculations are restricted to the two nuclei, namely $^{173}$Yb and $^{163}$Dy, which were previously processed by Ermamatov et al \cite{9,15} in order to correct the results that have already been obtained with their erroneous formula on one hand, and to see the effect of DDM on these calculations on the other hand.   

\par The second one is the formalism proposed by Bonatsos et al. \cite{5}, allowing the mass to depend on the nuclear deformation. This approach has been firstly achieved by applying  Davidson \cite{5} and Kratzer \cite{6} potentials to a huge number of $\gamma$-unstable and axially symmetric prolate deformed nuclei. It has also been extended to a conjunction between the prolate $\gamma$-rigid and $\gamma$-stable collective motions \cite{Chabab16}, and then tested on  Davydov-Chaban Hamitonian to describe triaxial shape nuclei around $\gamma=\pi/6$ \cite{Chabab18}.

%\par The third one is a new formalism which  has  introduced the minimal length in nuclear structure \cite{Chabab16B}. This formula is inspired from  Heisenberg algebra and  modifies the momentum operator according to the requirements of the  Generalized Uncertainty Principle (GUP). Afterwards, this approach was used in Ref. \cite{Chabab18B} to study the collective motion in prolate $\gamma$-rigid nuclei with a scaled Davidson potential in $\beta$ shape variable by means of a Quantum Perturbation Method (QPM).  Moreover, in Ref. \cite{Chabab19} a systematical  comparison between Coulomb and Hulthèn potentials within this formalism has been performed in transitional nuclei near the critical point symmetry X(3).% 
\par An investigation similar to the different mass parameters  approach of Ref. \cite{71} has been extended in Ref. \cite{10} for the collective single-particle ground state properties  of  deformed odd-A nuclei such as $^{163,165}$Er \cite{Ermamatov12} and $^{173}$Yb \cite{9}. This latter approach has been improved in Ref. \cite{15} by adding the Coriolis interaction between the rotational and single-particle motion of the  odd nucleon. Its influence was tested by applications on the experimental data of $^{163}$Dy and $^{173}$Yb \cite{15}, while in the present study we consider the projection of the nuclear total angular momentum onto the third axis $K$ and that of the external nucleon $\Omega$ as conserved quantities (i.e. $K$ and $\Omega$ are good quantum numbers), which means that the Coriolis interaction does not make a contribution. 

%However, the Coriolis effects are devoted in several works, such as in Refs. \cite{Minkov07}, coherent quadrupole-octupole oscillations and rotations with the Corilios coupling between the even-even core and the unpaired nucleon was applied  to odd nuclei within the deformation shell model (DSM). Also, in Ref. \cite{Minkov10} the Coriolis decoupling strength in nuclear single-particle (s.p.) states with mixed parity was examined within a reflection-asymmetric deformed shell model.
\par The purpose of the present work is to investigate a new  generalized version of the collective quadrupole Bohr Hamiltonian with different deformation-dependent  mass parameters, firstly developed in \cite{14}.  We will then propose a combination of the first and second approach mentioned above. Here, the Davidson and harmonic oscillator potentials are taken to  characterize  the $\beta$ and $\gamma$ vibrations, respectively. This choice is dictated by the number of interesting works that have been achieved with these potentials \cite{9,14,5,15,Yigitoglu}, which will allow a  comparison of our analytical results to those obtained by other authors. 

The analytical expressions of spectra, wave functions and reduced E2 transition probabilities are obtained by means of the asymptotic iteration method (AIM) \cite{11}. Therefore, we test this extended model to $^{173}$Yb and $^{163}$Dy  nuclei, by evaluating the experimental observables like energy spectra and B(E2) transition probabilities. Besides, we will study the effect of the deformation mass parameter which is used either : when we consider three different mass parameters or when  only a global mass parameter.As mentioned in \cite{9}, the  properties of the ground states of odd-mass nuclei that have an angular momentum $\ge 5/2$ can be studied with this model.    
\par This paper is organized as follows : In Section 2 we present the Bohr Hamiltonian with mass coefficients for the case of odd-mass nucleus, that we use in Section 3 in accordance with deformation-dependent mass formalism. Analytical expressions for the energy levels and excited-state wave functions of the model are presented in Section 4, while the B(E2) transition probabilities are given in Section 5. The numerical results for energy
spectra and B(E2) are presented, discussed, and compared with experimental data in Section 6, while Section 7 is devoted to our conclusions. The formulas of special cases of energy spectrum are given in Appendix A, while Appendix B is  dedicated to collect the used formulas for the calculations of B(E2).

\section{ Bohr Hamiltonian with mass parameters } 
According to the approach of Jolos and von Brentano \cite{71} for an odd-mass nucleus, for small harmonic  $\beta$ and $\gamma$ oscillations  of a deformed nuclear surface with respect to the equilibrium values $\beta_0 \neq 0$ and $\gamma_0 \approx 0$, the corresponding  Hamiltonian with three different mass parameters, can be written as 
   \begin{equation}
H=H_{\text{vib}}+H_{\text{rot}}+H_{\text{int}},
\label{diseqn6}
\end{equation}
where the operator describing the $\beta$ and $\gamma$ vibrations of the nuclear core surface  is 
\begin{multline}
		H_{\text{vib}}=-\frac{\hbar^2}{2}\Bigg\{\frac{1}{B_{ \beta}}\frac{\partial^2}{\partial\beta^2}+\frac{2}{B_{ \gamma}}\frac{1}{ \beta}\frac{\partial}{\partial\beta} +\frac{2}{B_{ \beta}}\frac{1}{ \beta}\frac{\partial}{\partial\beta}+\frac{1}{B_{ \gamma}\beta^2}\frac{1}{ \gamma}\frac{\partial}{\partial\gamma}\Bigg(\gamma\frac{\partial}{\partial\gamma}\Bigg)
		-\frac{1}{B_{ \gamma}}\frac{1}{4\beta^2}\\ \times\Bigg(\frac{1}{\gamma^2}+\frac{1}{3}\Bigg)\Big(L_{3}-j_{3}\Big)^2\Bigg\}+V(\beta,\gamma),
		%\nonumber
		\label{diseqn7}
	\end{multline}
The nuclear rotational energy operator  can conveniently be represented as 
\begin{equation}
H_{\text{rot}}=\frac{\hbar ^2}{6B_{\text{rot}}\beta^2}\Big(L^2+j^2-L_{3}^2-j_{3}^2\Big),
\label{diseqn8}
\end{equation}
and the interaction operator which takes into account non-spherical part of the field of the core  is given by the expression
	\begin{equation}
			H_{\text{int}}=-\beta_0 \langle T(r) \rangle \Big(3j_{3}^2-j^2\Big),
		%\nonumber
		\label{diseqn9}
	\end{equation} 
where $B_{\text{rot}}$, $B_{\beta}$ and $B_{\gamma}$ are three different mass coefficients for rotational, $\beta$-, $\gamma$- motion, respectively. $L$ is the total angular momentum, where $L_3$ is the eigenvalue of the projection of angular momentum on the principal axis of nucleus, $j$ and $j_3$ are the angular momentum operator of a single nucleon, and its projection. $\beta_0$ is the equilibrium value of the nuclear surface $\beta$-oscillations and $T(r)$  is a function of the distance between the single
nucleon and the center of the nuclear core, while $\langle T \rangle$ is its  average value  over internal states of the external nucleon and zero nuclear surface oscillations \cite{4,23}. 
%Here, we have omitted the Coriolis contribution $2(L_{1}j_{1}+L_{2}j_{2})$ , which is not
%diagonal, and because j ,K, and $\Omega$ are good quantum numbers,
%only diagonal elements of the Hamiltonian contribute to the
%energy;\\
%
%In Refs. \cite{23} and \cite{4}, T (r) is a function of the distance
%between the single nucleon and the center of the core nucleus.
%The values of both K and $\Omega$ should be up to j , and satisfy
%the condition $ K-\Omega$= 2m \cite{23}, where m is an integer.
\section{Connection between Deformation-Dependent mass and different mass parameters }
By following the procedure in Ref. \cite{7}, we wish to construct a Bohr equation with three deformation dependent  mass coefficients, in accordance with the DDM formalism described in Ref. \cite{5}. So, the mass tensor of the collective Hamiltonian becomes 

\begin{equation} 
B=\frac{\langle i|B_0|i \rangle}{(f(\beta))^2},
\label{diseqn10}
\end{equation}
where $i=${ g.s.}, $\beta$ or $\gamma$ (g.s. is often replaced by $rot$) corresponding to three separable state bands of nuclei, namely : the ground state band, the $\beta$ and $\gamma$ vibrational bands, each one of these will have its own mass coefficient equal to its average value over the wave function of the considered state, such as : $\langle g.s.|B_0|g.s. \rangle \equiv B_{\text{rot}}$, $\langle \gamma|B_0|\gamma \rangle \equiv B_{\gamma}$ and $\langle \beta|B_0|\beta \rangle \equiv B_{\beta}$ defined for each band. $f$ is the deformation function depending only on the radial coordinate $\beta$. Therefore, only the $\beta$ part of the resulting equation will be affected. 
\par The explicit equation  reads as \cite{14}
\begin{multline} 
\frac{\hbar ^2}{2\langle i|B_{0}|i \rangle}$\Big($ -\frac{\sqrt{f}}{\beta^4}\frac{\partial}{\partial\beta} {\beta^4f}\frac{\partial}{\partial\beta}\sqrt{f}- \frac{f^2}{\beta^2\sin3\gamma}\frac{\partial}{\partial\gamma}\sin3\gamma\frac{\partial}{\partial\gamma}\\+
\frac{f^2}{4\beta^2}\sum_ {k=1,2,3}\frac{(Q_{k}-j_{k})^{2}}{\sin^2(\gamma-\frac{2}{3}\pi k)} \Big)\Psi -f^2\beta\langle T \rangle(3j_{3}^2-j^2)\Psi+V_{eff}\Psi=E\Psi,
\label{diseqn14}
\end{multline}
with,
\begin{equation}
V_{eff}=V(\beta,\gamma)+\frac{\hbar^2}{2\langle i|B_{0}|i \rangle}\Big(\frac{1}{2}(1-\delta-\lambda)f\bigtriangledown^2f
+(\frac{1}{2}-\delta)(\frac{1}{2}-\lambda)(\bigtriangledown f)^{2}\Big),
    \label{diseqn15}
\end{equation}
where $\delta$ and $\lambda$ are free parameters origenated from the construction procedure of the kinetic energy term within DDM formalism.
\section{Energy spectrum and excited-state wave functions}
\par Exact separation of variables $\beta$ and $\gamma$ can be achieved for potentials  using the convenient form \cite{14} $V(\beta,\gamma)=U(\beta)+f^2W(\gamma)/\beta^2$, in which the potential depending only on $\gamma$  has a minimum around $\gamma=0$. In the same context, the total wave function can be constructed as 
\begin{equation}
\Psi=F(\beta)\chi(\gamma)|LMjKm\rangle,
%\Psi=F(\beta)\chi_{n_{\gamma}|m|}(\gamma)|LMjKm\rangle
 \label{diseqn17}
\end{equation} 
where the rotational wave function $|LMjKm\rangle $ has been expanded \cite{15,4} in terms of the Wigner $D(\theta_i)$-function of the Euler angles, and $\varphi(x_i)$ the eigenfunction of the single-particle states, in the following form    
\begin{equation}
|LMjKm\rangle=\sqrt{\frac{2L+1}{16\pi^2}}\Big[D^L_{MK}(\theta_{i})\varphi^j_{K-2m}(x_{i})
+(-1)^{L-j}D^L_{M-K}(\theta_{i})\varphi^j_{-K+2m}(x_{i})\Big],
\label{diseqn18}
\end{equation}
where $K$  the projection of the nuclear angular  momentum on the third axis  connected with the nucleus and $\Omega=K-2m$ represents  the projection of the angular momentum of the external nucleon on the same axis \cite{4}. Note that $m$ should be an integer. As a result, Eq. (\ref{diseqn14}) can be  separated into three equations 
\begin{multline}
\Bigg[\frac{\hbar ^2}{2\langle i|B_{0}|i \rangle}$\Big($ -\frac{\sqrt{f}}{\beta^4}\frac{\partial}{\partial\beta} {\beta^4f}\frac{\partial}{\partial\beta}\sqrt{f}+\frac{f^2}{\beta^2}\Lambda+\frac{1}{2}(1-\delta-\lambda)f\bigtriangledown^2f
+(\frac{1}{2}-\delta)(\frac{1}{2}-\lambda)(\bigtriangledown f)^{2}\Big)\\+V(\beta)\Bigg]F(\beta)=EF(\beta),
\label{diseqn19}
\end{multline}
\begin{multline}
  \Bigg[ -
   \frac{\hbar^2}{2B_{\gamma}}\Bigg( \frac{1}{\sin3\gamma}\frac{\partial}{\partial\gamma}\sin3\gamma\frac{\partial}{\partial\gamma}-\frac{1}{4}\bigg(\frac{1}{\gamma^2}+\frac{1}{3}\bigg)\Big(L_{3}-j_{3}\Big)^2\Bigg)\\+W(\gamma)\Bigg]\chi_{n_{\gamma}|m|}(\gamma)=\Lambda'\chi_{n_{\gamma}|m|}(\gamma),  
  \label{diseqn20}
\end{multline}
%\begin{multline}
%\Big[ -
%\frac{\hbar^2}{2B_{\gamma}}\Big( \frac{1}{\sin3\gamma}\frac{\partial}{\partial\gamma}\sin3\gamma\frac{\partial}{\partial\gamma}-\frac{1}{4}(\frac{1}{\gamma^2}+\frac{1}{3})(L_{3}-j_{3})^2+\Lambda\Big)\\+W(\gamma)\Big]\chi_{K|m|}(\gamma)=\Lambda'\chi_{K|m|}(\gamma)  
%\label{diseqn}
%\end{multline}
\begin{equation}
\Bigg[\frac{\hbar^2}{6B_{rot}}\Big(L^2+j^2-L^2_{3}-j^2_{3}\Big)-\beta^3 \langle T \rangle \Big(3j_{3}^2-j^2\Big)\Bigg]|LMjKm\rangle=\bar{\Lambda}|LMjKm\rangle,  
\label{diseqn21}
\end{equation}
where $\Lambda=\Lambda'+2\bar{\Lambda}$ is the parameter coming from the exact separation of variables, $\Lambda'$ is the eigenvalues corresponding to $\gamma$-vibrations, while  $\bar{\Lambda}$ is the internal state energy corresponding to the rotational energy operator $H_{\text{rot}}$ and the interaction operator $H_{\text{int}}$. It is represented by the term 
%%% for each value value of the total angular momentum of the system described by the quantum number $L=\frac{1}{2}, |frac{3}{2}$ etc
\begin{multline}
\frac{B_{\beta}}{\hbar^2} \bar{\Lambda}=\frac{B_{\beta}}{6B_{rot}}\Big(L(L+1)+j(j+1)-K^2-(K-2m)^2\Big)-\frac{1}{6 \xi} \Big(3(K-2m)^2-j(j+1)\Big),
\label{diseqn22}
\end{multline}
with  $\xi=\frac{\hbar^2}{6B_{\beta}\beta^3\langle T\rangle}$. Concerning  the $\gamma$-angular part  Eq. \eqref{diseqn20}, the potential $W(\gamma)$ is assumed to be harmonic oscillator around $\gamma=0$ as in Ref. \cite{14}, namely  $W(\gamma)=\frac{1}{2}(\beta^4_{0}C_{\gamma})\gamma^2$, where $C_{\gamma}$ is a free parameter. It can be seen that Eq.\eqref{diseqn20} is similar to the $\gamma$ part of the differential equation in Ref. \cite{14}.  The corresponding analytical solution of Eq. \eqref{diseqn20}  is given with eigenvalues \cite{14}
 \begin{equation}
\frac{B_{\beta}}{\hbar^2}\Lambda'=\frac{2}{g}\frac{B_{\beta}}{B_{\gamma}}\Big(1+n_{\gamma}\Big)+\frac{m^2}{3}\frac{B_{\beta}}{B_{\gamma}},\label{diseqn23}
\end{equation}
and the corresponding eigenfunctions,  are obtained in terms of the Laguerre polynomials as 
 \begin{equation}
 \chi_{n_{\gamma}|m|}(\gamma)=N_{n_{\gamma},|m|}\gamma^{|m|}e^{\dfrac{-\gamma^2}{2g}}L^{|m|}_{{\tilde{n}_{\gamma}}}\Big(\frac{\gamma^2}{g}\Big),
 \label{diseqn31}
 \end{equation}
 with $g =\frac{1}{\beta^2_{0}}\frac{\hbar}{\sqrt{B_{\gamma}C_{\gamma}}}$.  ${\tilde{n}_{\gamma}}=\frac{n_{\gamma}-|m|}{2}$,  $n_{\gamma}$ is the quantum number related to $\gamma$-oscillations and $N_{n_{\gamma},|m|}$ is a normalization constant determined from the normalization condition. This leads to  
\begin{equation}
N_{n_{\gamma},|m|}=\Big(\frac{2}{3}g^{-1-|m|}\dfrac{{\tilde{n}_{\gamma}}!}{\Gamma(|m|+{\tilde{n}_{\gamma}}+1)}\Big)^{\frac{1}{2}},
\label{diseqn33}
\end{equation}
Concerning the $\beta$-oscillation states of deformed odd-A nuclei, they are determined by the solution of radial equation \eqref{diseqn19} with Davidson potential, $V(\beta)=V_0\Big(\frac{\beta}{\beta_0}-\frac{\beta_0}{\beta}\Big)^2$, where $V_0$ represents the depth of the minimum, located at $\beta_0$ and $f(\beta)$ the deformation function.  According to specific form of Davidson potential we are going to consider for the deformation function the special form $f(\beta)=1+a\beta^2$, with $a<<1$. In fact, for each potential an appropriate deformation function will be used. With these last considerations, the resulting radial equation becomes the same as the $\beta$ part of the differential equation given in Sect. VI of Ref. \cite{14}. Note that the only difference between them resides in  $\Lambda$ the eigenvalue of the exact separation of variables, depending on the nature of the nucleus. \\Thus, the energy spectrum of the radial equation is determined by the following expression,  \cite{14}
\begin{multline}
E_{n_{\beta}n_{\gamma}L|m|}=\frac{\hbar^2}{2B_{\beta}}\Big(k_{0}+\frac{a}{2}(2+\frac{B_{\beta}}{B_{\gamma}}+2p+2q+pq)+2a(2+p+q)n_{\beta}+4an^2_{\beta}\Big),
\label{diseqn46}
\end{multline}
where $n_{\beta}$ is the principal quantum number of $\beta$ vibrations, and 
\begin{equation}
\begin{gathered}
q\equiv  q_{n_{\gamma}}(L,|m|)=\sqrt{1+4k_{-2}}, \quad \quad \quad \\p\equiv p_{n_{\gamma}}(L,|m|)=\sqrt{4\frac{B_{\beta}}{B_{\gamma}}-3+4\frac{k_{2}}{a^2}},
\end{gathered}
\label{diseqn47}
\end{equation}
\begin{align}
k_2=&\frac{a^2}{2}\Big[\Big(1+\frac{B_{\beta}}{B_{\gamma}}\Big)\Big(6\frac{B_{\beta}}{B_{\gamma}}+(1-2\delta)(1-2\lambda)+5(1-\delta-\lambda)\Big)+\frac{2B_{\beta}}{\hbar^2}\Lambda\Big]+\frac{2g_{\beta}}{\beta^4_{0}},&\nonumber
\\
k_0=&\frac{a}{2}\Big[\Big(1+\frac{B_{\beta}}{B_{\gamma}}\Big)\Big(8\frac{B_{\beta}}{B_{\gamma}}+5(1-\delta-\lambda)\Big) +\frac{4B_{\beta}}{\hbar^2}\Lambda\Big]-\frac{4g_{\beta}}{\beta^2_{0}},&\label{diseqn42}\\
k_{-2}=&\frac{B_{\beta}}{B_{\gamma}}\Big(1+\frac{B_{\beta}}{B_{\gamma}}\Big)+\frac{B_{\beta}}{\hbar^2}\Lambda+2g_{\beta},&
  \nonumber
\end{align}
where $g_{\beta}=\frac{B_{\beta}V_{0}\beta^2_{0}}{\hbar^2}$, the excitation energies \eqref{diseqn46} do not depend on $j$ because $j$,$\Omega$,and  $K$  are conserved \cite{9},so they depend on five quantum numbers,namely: $n_{\beta}$, $n_{\gamma}$, $L$, $K$ and $m$, and nine parameters : $g$, $g_{\beta}$, $\xi$, $B_{\beta}/B_{\gamma}$, $B_{\beta}/B_{\text{rot}}$ ratio of the mass coefficients, $a$ the deformation mass parameter, $\beta_0$ the minimum of the potential and the free parameters $\delta$ and $\lambda$ coming from the DDM formalism. In the numerical results Section, a comparison to the experiment will be carried out by fitting the theoretical spectra to experimental data. Finally, it will be shown that the predicted energy levels turn out to be independent of the choice made for $\delta$ and $\lambda$.
\par In the limit cases of the energy spectrum, our general formula \eqref{diseqn46} can well reproduce three special cases, namely : the first without mass coefficients i.e. if we assume $B_{\beta}=B_{\gamma}=B_{\text{rot}}=1$, the second in the limit of no dependence of the mass on the deformation, i.e. $a=0$ and the third standard case, when $B_{\beta}$=$B_{\gamma}$=$B_{\text{rot}}$ and $a=0$. All this special cases are carried out in the  Appendix \ref{appendixA} with their energy spectrum expressions.
 
\par The relevant radial eigenfunctions of Eq. \eqref{diseqn19} are found in Ref. \cite{14} to be 
\begin{align}
R(t)=&N_{n_{\beta}}2^{-(1+\frac{B_{\beta}}{B_{\gamma}})/2-(q+p)/4}a^{-(1+q)/4}(1-t)^{(1+2\frac{B_{\beta}}{B_{\gamma}}+p)/4}(1+t)^{(q+1)/4}P^{(q/2,p/2)}_{n_{\beta}}(t),& \nonumber \\
t=&\frac{-1+a\beta^2}{1+a\beta^2}.& 
\label{diseqn60}
\end{align}
  $P_n^{(\alpha,\beta)}(t)$  denotes the Jacobi polynomials \cite{19}, while the normalization coefficient $N_{n_{\beta}}$ is given by 
\begin{align}
N&_{n_{\beta}}=\Big(2a^{q/2+1}n_{\beta}!\Big)^{\frac{1}{2}}\Bigg[\frac{\Gamma\big(n_{\beta}+\frac{q+p}{2}+1\big)\Gamma\big(2n_{\beta}+\frac{q+p}{2}+1+\frac{B_{\beta}}{B_{\gamma}}\big)}{\Gamma\big(n_{\beta}+\frac{q}{2}+1\big)\Gamma\big(n_{\beta}+\frac{B_{\beta}}{B_{\gamma}}+\frac{p}{2}\big)\Gamma\big(2n_{\beta}+\frac{q+p}{2}+1\big)}\Bigg]^{\frac{1}{2}}.
\label{diseqn61}
\end{align}
The reduction of the present wave functions Eq. \eqref{diseqn60} and the normalization constant Eq. \eqref{diseqn61} to the form they have in $B_{\beta}$=$B _{\gamma}$=$B _{\text{rot}}$= 1 limit are in agreement with Eq. (108) and Eq. (112) of Ref. \cite{5}, respectively, when the latter are simplified to consider only even-mass nuclei and conserved $K$. On the other hand, in the special case of no dependence of the mass on the deformation $a \rightarrow$ 0, the excited-state wave functions are found in Ref \cite{14} to be 
\begin{equation}
R(\beta)=N_{n_{\beta}}\beta^{\frac{1}{2}(1+q)}e^{-b\beta^2}L^{\frac{1}{2}q}_{n_{\beta}}(2b\beta^2),
\label{diseqn63}
\end{equation}
where $b=\sqrt{\frac{g_{\beta}}{2\beta^4_{0}}}$,  $L$ denotes the Laguerre polynomials and $N_{n_{\beta}}$ is a normalization coefficient reduced to the form 
\begin{equation}
N_{n_{\beta}}=\Bigg[2(2b)^{\frac{1}{2}q+1}\frac{n_{\beta}!}{\Gamma(n_{\beta}+\frac{1}{2}q+1)}\Bigg]^{\frac{1}{2}},
\label{diseqn67}
\end{equation}
\section{ B(E2) transition probabilities }
The B(E2) transition rates from an initial to a final state are given by \cite{20},
\begin{equation}
B(E2;L_{i}K_{i}\longrightarrow L_{f}K_{f})=\frac{5}{16\pi}\frac{|\langle L_{f}K_{f}||T^{(E2)}||L_{i}K_{i}\rangle|^2}{2L_{i}+1},
\label{diseqn68}
\end{equation}
and the reduced matrix element can be obtained by using
the Wigner-Eckrat theorem \cite{20},
\begin{equation}
\langle L_{f}M_{f}K_{f}|T^{(E2)}_{M}|L_{i}M_{i}K_{i}\rangle  \label{diseqn69}
=\frac{(L_{i}2L_{f}|M_{i}MM_{f})}{\sqrt{2L_{f}+1}}\langle L_{f}K_{f}||T^{(E2)}||L_{i}K_{i}\rangle.
\end{equation}
The final result \cite{21} reads
\begin{equation}
B(E2;n_{\beta}Ln_{\gamma}K|m|\longrightarrow n'_{\beta}L'n'_{\gamma}K'|m'|)\label{diseqn70}
=\frac{5}{16\pi}\langle L,K,2,K'-K|L',K'\rangle^2I^2_{n_{\beta}L,n'_{\beta}L'}C^2_{n_{\gamma},|m|,n'_{\gamma},|m'|},
\end{equation}
with
\begin{equation}
I_{n_{\beta}L,n'_{\beta}L'}=\int \beta F_{L,n_{\beta}}(\beta)F_{L',n'_{\beta}}(\beta) \beta^{2+2\frac{B_{\beta}}{B_{\gamma}}}\,\mathrm{d\beta}
=\int \beta R_{L,n_{\beta}}(\beta)R_{L',n'_{\beta}}(\beta) \,\mathrm{d\beta},
\label{diseqn71}
\end{equation}
where $C_{n_{\gamma},|m|,n'_{\gamma},|m'|}$ contains the integral over $\gamma$. For $\Delta m= 0$ corresponding to transitions (g.s.$\longrightarrow$ g.s.), ($\gamma \longrightarrow \gamma$), ($\beta  \longrightarrow \beta$) and
$(\beta \longrightarrow g.s.)$, the $\gamma$-integral part reduces to the orthonormality
condition of the $\gamma$-wave functions : $C_{n_{\gamma},|m|,n'_{\gamma},|m'|}=\delta_{n_{\gamma},n'_{\gamma}}\delta_{m,m'}$. For $|\Delta m| = 1$ corresponding to transitions
($\gamma \longrightarrow g.s.$), ( $\gamma \longrightarrow \beta$), this integral takes rather the form.
\begin{equation}
C_{n_{\gamma},|m|,n'_{\gamma},|m'|}=\int \sin \gamma\;\chi_{n_{\gamma}|m|}\chi_{n'_{\gamma}|m'|}|\sin 3\gamma|\mathrm{d\gamma}.
\label{diseqn72}
\end{equation}
 In the next sections, all values of B(E2) are calculated in units of $B(E2;\frac{9}{2}_{\text{g.s.}} \rightarrow \frac{5}{2}_{\text{g.s.}}).$
 \section{ Numerical results of energy and B(E2) ratios and discussion}
\par Before starting our calculations of energy spectra and transition rates for the two deformed $^{173}Yb$ and $^{163}Dy$ nuclei, in the special case  without DDM formalism (i.e. $a=0$), we have determined the optimal values  of   the free parameters $B_{\beta}/B_{\gamma}$, $B_{\beta}/B_{rot}$, $g$ and  $g_{\beta}$ by fitting the energy formula \eqref{diseqn51} (Appendix \ref{appendixA}) on the available experimental data, except the value of  $\xi$ that takes into account the interaction of the extra nucleon with the core, which is chosen as $\xi=1$, because, we have considered the quantum numbers $K$ and $\Omega$ are conserved. Then the parameter $\xi$ will not change the results  significantly. These parameters are adjusted to reproduce the experimental data by applying a least-squares fitting procedure for each considered nucleus. For this purpose we have minimized the root mean square (r.m.s) deviation between the theoretical values and the experimental data via the following quantity factor 
%\begin{longtable}[htb]
%\begin{center}
\begin{longtable}{cccccccccc}
 \hline \multicolumn{7}{c}{ nucleus}\quad \quad \quad \quad{$g$}\quad \quad \quad \quad$ {g_{\beta}}$\quad \quad \quad \quad $B_{\beta}/B_{\gamma}$ \quad \quad \quad \quad $B_{\beta}/B_{rot}$\quad \quad \quad \quad $g(B_{\beta}=B_{\gamma}=B_{rot})$\quad \quad \quad \quad $g_{\beta}(B_{\beta}=B_{\gamma}=B_{rot})$ \\ \hline 
$^{173}Yb$ \quad \quad \quad \quad   &0.0094\quad  &1033.26\quad \quad  &1.3008\quad \quad \quad \quad \quad   &7.5\quad \quad \quad \quad \quad \quad \quad &0.0464\quad \quad \quad \quad \quad  &0.3643 \\ \hline 
$^{163}Dy$\quad \quad \quad \quad    &0.003\quad  &235.44\quad \quad  &2.3731\quad \quad \quad \quad \quad   &7.9063\quad \quad \quad \quad \quad \quad \quad &0.0628\quad \quad \quad \quad \quad  &0.5853 \\ \hline 
%\end{longtable}
%\end{center}
\caption{The values of free parameters fitted to experimental data by using Eqs. \eqref{diseqn51} and  \eqref{diseqn54} in the case where  $B_{\beta}\neq B_{\gamma}\neq B_{\text{rot}}$ and $B_{\beta}= B_{\gamma}= B_{\text{rot}}$, respectively.}
\label{tab1}
\end{longtable}

\begin{equation}
	\sigma=\sqrt{\frac{\sum_{i=1}^n(E_i(exp)-E_i(th))^2}{(n-1)E(7/2_{\text{g.s.}})^2}},
	\label{diseqn73}
	\end{equation}
	where $n$ denotes the number of considered levels, $E_i(exp)$ and $E_i^{th}$ represent the experimental and theoretical energies of the $i$-th level, respectively.   $E(7/2_{\text{g.s.}})$ is the g.s. band head energy.
	\par All bands (i.e. ground state, $\beta$ and $\gamma$)  are labelled by the quantum numbers $n_{\beta}$, $n_{\gamma}$, $m$, $L^{\pi}$, $K$ and $j$, such as the ground state band (g.s.) is characterized by $n_{\beta}$=0, $n_{\gamma}$ = 0,$m$=0 , the $\beta$-band by $n_{\beta}=1$,$n_{\gamma} = 0$, $m$=0 , and the $\gamma$-band by $n_{\beta}$=0, $n_{\gamma}$ = 1, $m$=1, while the  appropriate value of the angular momentum of the external nucleon is $j=7/2$. In fact, the  shell-model calculations achieved in Ref. \cite{15} predict that the major contribution to the ground state structure of  considered nuclei comes from the neutron $f_{7/2}$ orbital.  The value of $K$ is fixed to $5/2$ for both nuclei in accord with \cite{9}.

	\par The obtained optimal parameters for the two considered nuclei are given in Table 1 in both cases, namely : the case where the mass parameters  are different and the case where they are equal within and outside the DDM formalism. Together with the energy formula Eq. \eqref{diseqn51} (Appendix \ref{appendixA}) and the obtained parameters, we have evaluated the energy spectra of the considered nuclei as well as the corresponding transition rates. But, here we have to bear in mind that our formula Eq. \eqref{diseqn51} (Appendix \ref{appendixA}) is a correction to the one previously obtained by Ermamatov et al. \cite{9}. So, the obtained values outside the DDM formalism ($a=0$) are just the results which should be obtained by the authors \cite{9}. 
	\par The DDM formalism has been introduced in the present work  in order to see its impact in both above-cited cases. For that, we recalculated the energy ratios with the more elaborated formula given in Eq. \eqref{diseqn46}. Such an expression contains two  supplementary parameters, namely $a$ and $\beta_0$. The optimal values of both parameters are evaluated through r.m.s fits of energy levels by making use of Eq. \eqref{diseqn73}  for
each band of each nucleus. The obtained values are summarized in Table 2 and Table 3 for $^{173}Yb$ and $^{163}Dy$ nuclei, respectively.
	\par From Table 4, where the energy ratios  $E(L_{\text{g}.s.})/E(7/2^+_{\text{g}.s.})$, $E(L_{\beta})/E(7/2^+_{\text{g}.s.})$ and $E(L_{\gamma})/E(7/2^+_{\text{g}.s.})$ of $^{173}$Yb are presented, one can see that outside the DDM formalism ($a=0$), the obtained results with different mass parameters are more precise than those obtained with equal ones. Also, the introduction of the DDM parameter ($a\neq 0$) has improved the obtained results in both cases, namely : $B_{\beta} \neq B_{\gamma} \neq B_{rot}$ and $B_{\beta}=B_{\gamma}=B_{rot}$ but with keeping the prevalence of the first one. From the same Table, one can observe  that all  level bands are more sensitive to the effect of the DDM parameter $a$ particularly in the case of equal mass parameters, while when the mass parameters are different, the $\gamma$-band has not been influenced by the DDM parameter.
	\par In Table 5, we present the calculated energy ratios normalized to the first $g.s$ excited level $\frac{7}{2}^-$ for $^{163}$Dy in the $g.s$ and $\beta$-bands only because the  experimental data are not available in the $\gamma$-band. Here again, the obtained results with $B_{\beta} \neq B_{\gamma} \neq B_{rot}$ are better than those corresponding to $B_{\beta} = B_{\gamma} = B_{rot}$ and the effect of the DDM formalism is more pronounced in the later case.
%	\par In Table 7, are shown the $g.s$, $\beta$ and $\gamma$ bands levels normalized to the first $g.s$ excited level $\frac{7}{2}^-$. Here, one can see that outside the DDM formalism $(a=0)$ the obtained values in the case of $B_{\beta} \neq B_{\gamma} \neq B_{rot} $ are so better than the obtained ones in the case of equal mass parameters in all bands. However, in the presence of the DDM parameter $(a \neq 0)$ the calculated energy ratios for $B_{\beta}=B_{\gamma}=B_{rot}$ were much improved and become comparable to those obtained with $B_{\beta} \neq B_{\gamma} \neq B_{rot}$.

 \par Similarly, we have calculated the intraband $B(E2;L'_{\text{g.s.}} \rightarrow L_{\text{g.s.}})$ and interband $B(E2;L_{\beta} \rightarrow L_{\text{g.s.}})$ and $B(E2;L_{\gamma} \rightarrow L_{\text{g.s.}})$ transition probabilities given in Eqs.  \eqref{B1}, \eqref{B2} and \eqref{B3} (Appendix \ref{appendixb}), respectively, normalized to transition rate from the first excited level in the g.s. band  $B(E2;9/2_{\text{g.s.}} \rightarrow 5/2_{\text{g.s.}})$  for the same nuclei in both cases $B_{\beta} \neq B_{\gamma} \neq B_{\text{rot}} $ and $B_{\beta} = B_{\gamma} = B_{\text{rot}} $ within and without the DDM formalism. For each nucleus, the parameters obtained by fitting the spectra have been used. The results are shown in Table 6 and 7. Only theoretical calculations of interband transitions rates from the $\beta$ and $\gamma$ bands to g.s. band are presented  since there are no experimental values for them as yet. Concerning the intraband transitions within the g.s. band, it is clearly shown that our results in the case of $B_{\beta} \neq B_{\gamma} \neq B_{\text{rot}} $ are better than those with  $B_{\beta} = B_{\gamma} = B_{\text{rot}} $ and at higher spins the intraband E2 transition probabilities  are affected by these differences of mass parameters and showed an overall agreement with the experimental data. However, one can remark that the deformation mass parameter has no effect in the case of different mass coefficients, while for equal mass parameters, its effect is pronounced particularly for interband transitions.
 	\begingroup
\begin{figure*}
\begin{center}
\includegraphics[scale=0.9]{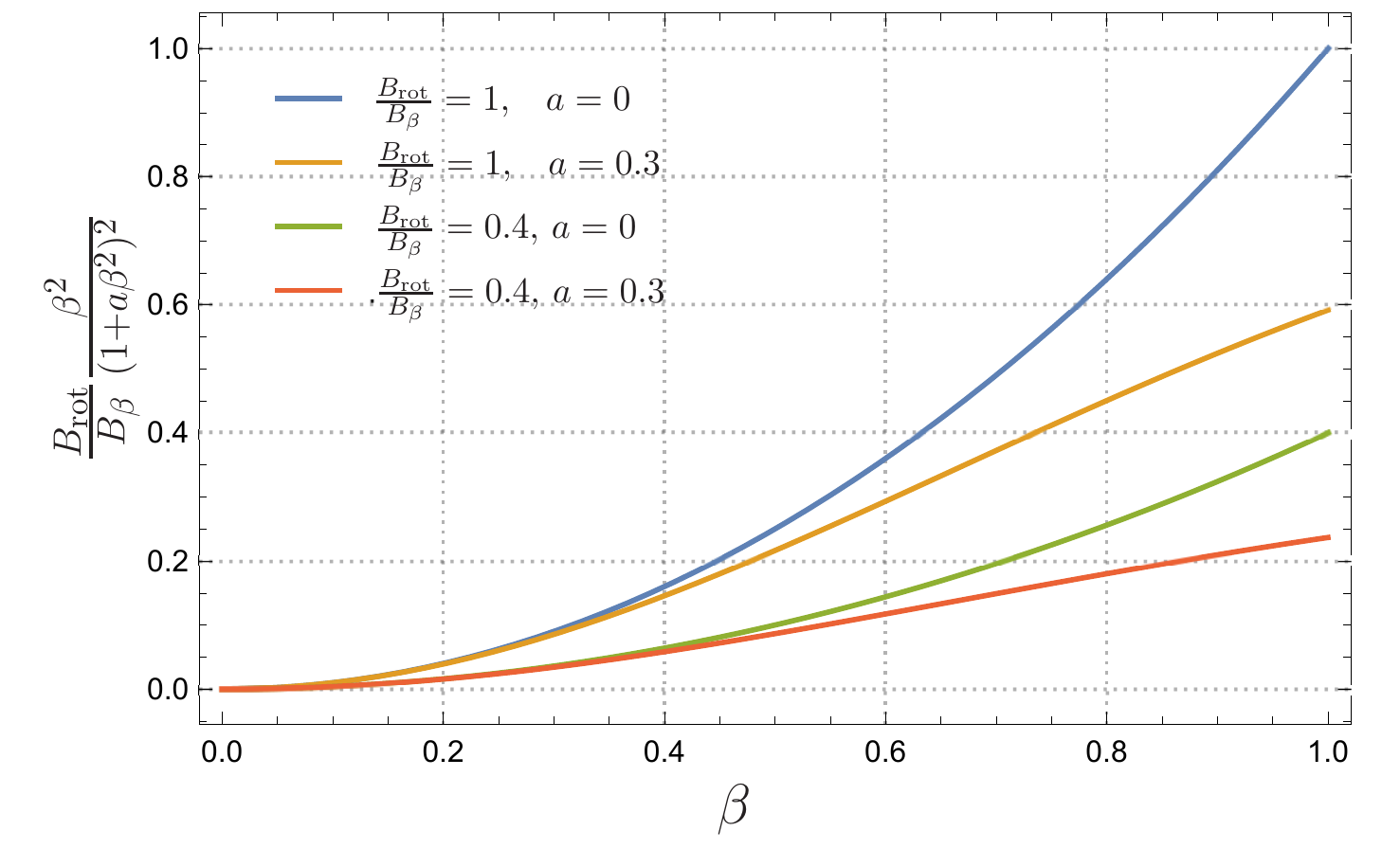}
\caption{ (Color online) The function $B_{\text{rot}}\beta^2/B_{\beta}(1+a\beta^2)^2$, to which moments of inertia are proportional, plotted as a function of the nuclear deformation $\beta$ in both cases $ B_{\text{rot}}\neq B_{\beta} $ and $ B_{\text{rot}}= B_{\beta} $ within (i.e. $a\neq 0$) and without (i.e. $a= 0$) the DDM formalism. }\label{fig1}
\end{center}
\end{figure*}
\endgroup

\par In addition to the above obtained results, here we have to notice the double effect of both formalisms, namely: the DDM and the different mass parameters on the variation of the moment of inertia. It is clear that in the case 	of the Bohr Hamiltonian with different deformation-dependent mass parameters as seen from Eq. \eqref{diseqn14}, the moment of inertia is defined by $(B_{\text{rot}}\beta^2/B_{\beta}f(\beta)^2)\sin^2(\gamma-2\pi k/3)$. The effect of the function  $B_{\text{rot}}\beta^2/B_{\beta}(1+a\beta^2)^2$ on the moment of inertia is shown in Fig. 1 for $ B_{\text{rot}}\neq B_{\beta} $ and $ B_{\text{rot}}= B_{\beta} $ within (i.e. $a\neq 0$) and without (i.e. $a= 0$) the DDM formalism. It is apparent that the increase of the moment of inertia is slowed down by the function of deformation $f(\beta)$ (i.e. when $a \neq 0)$) and even more in the presence of  different mass coefficients (i.e. when $ B_{\text{rot}}\neq B_{\beta} $). Then, the present approach, reduces significantly the rate of increase of the moment of inertia, removing a main important drawback \cite{32} of the model. 

\section{ CONCLUSION}
In this paper, we have studied two deformed odd-mass nuclei, $^{173}$Yb and  $^{163}$Dy, in the framework of the Bohr Hamiltonian with deformation-dependent mass coefficients using Davidson potential in $\beta$ shape and the harmonic oscillator in $\gamma$ potential. Analytical expressions have been obtained for the excited-state energies, wave functions and E2 transition probabilities. Such formulas are the corrected ones for those previously obtained in \cite{9}. Energy levels of the g.s., $\beta$ and $\gamma$ bands with $K=5/2$  as well as the interband transitions rates from the $\beta$ and $\gamma$ bands to g.s. band and intraband transitions within g.s. band, were calculated in both cases  $B_{\beta} \neq B_{\gamma} \neq B_{\text{rot}} $ and  $B_{\beta} = B_{\gamma} = B_{\text{rot}} $ within and without the DDM formalism and compared to available experimental data. Some predictions for transition rates are made where the experimental data are not available. Also, we have studied the effect of the deformation mass parameter on energy spectra and transition rates in both above-cited cases.  Moreover, we have shown the importance of the mass parameter to be introduced in numerical calculations, unlike what has been done by other authors who have neglected the important role played by this parameter in such calculations.
	
%\begin{longtable}[H]
%\centering
\newpage
\begin{longtable}{ccccccc}
%%\toprule
\hline \hline \hline
\qquad \qquad \qquad 
 $B_{\beta}\neq B_{\gamma}\neq B_{rot} $ && & & &  $B_{\beta}= B_{\gamma}= B_{rot} $\\
\hline
  &  & $a$=0\quad \quad   & DDM \quad \quad \quad \quad \quad \quad \quad & $a$=0& DDM &  \\ \hline \hline
%\midrule
 g.s \\
%25/2^- & & 22.01\quad \quad & 22.10\quad \quad \quad \quad \quad \quad \quad & 18.92& 22.16 & 21.63 \\
%27/2^- &25.71 &25.56\quad \quad & 25.68\quad \quad \quad \quad \quad \quad \quad & 21.56& 25.81 & 28.52 
$\sigma$ &  &  0.0399 & 0.0253\quad \quad \quad \quad \quad \quad \quad &1.2419 & 0.0235 &\\

a &  & \quad \quad & 0.001\quad \quad \quad \quad \quad \quad \quad & & 0.0881 & \\
$\beta_{0}$ &  & \quad \quad & 6.94\quad \quad \quad \quad \quad \quad \quad & & 1.029 & \\\hline \hline
 $\beta$ \\ 
%27/2 &0 &1\quad \quad & 5\quad \quad \quad \quad \quad \quad \quad & 0.00& 2 & 6 \\ \hline
$\sigma$ &  & 0.6839 & 0.6837\quad \quad \quad \quad \quad \quad \quad &1.1682 & 1.1682 & \\ 

a &  & \quad \quad & 0.1724\quad \quad \quad \quad \quad \quad \quad & & $1.94.10^{-11} $& \\
$\beta_{0}$ &  & \quad \quad & 0.0913\quad \quad \quad \quad \quad \quad \quad & & 3.8091 & \\\hline \hline
 $\gamma$ \\ 
%$21/2^-$ & 30.44 & 31.27\quad \quad  & 31.27\quad \quad \quad \quad \quad \quad \quad & 28.77& 29.09 &  \\
%23/2^{-} & 32.31 & 34.21\quad \quad  & 34.21\quad \quad \quad \quad \quad \quad \quad & 30.72& 31.09 & 

%25/2 &0 & 1\quad \quad & 2\quad \quad \quad \quad \quad \quad \quad & 0.00& 2 & 6 \\
%27/2 &0 &1\quad \quad & 5\quad \quad \quad \quad \quad \quad \quad & 0.00& 2 & 6 \\ \hline
$\sigma$ & & 0.347 & 0.347\quad \quad \quad \quad \quad \quad \quad &1.01& 0.9828 & \\ 

a &  & \quad \quad & $1.14.10^{-7}$\quad \quad \quad \quad \quad \quad \quad & & 0.00064 & \\
$\beta_{0}$ &  & \quad \quad & 0.3703\quad \quad \quad \quad \quad \quad \quad & & 1.8036 & \\\hline

 \hline \hline
 $\sigma_{total}$ & & 0.4135 & 0.4134\quad \quad \quad \quad \quad \quad \quad &1.0974& 1.0009 & \\ 
 a &  & \quad \quad & 0.0099\quad \quad \quad \quad \quad \quad \quad & & 0.0037 & \\
 $\beta_{0}$ &  & \quad \quad &0.3251\quad \quad \quad \quad \quad \quad \quad & & 1.3531 &\\\hline
 
 \hline \hline
%\bottomrule
%\end{longtable}
\caption { $\beta_{0}$ and  $a$  the position of the minimum of Davidson potential  and the deformation dependence of the mass parameter  respectively for the $^{173}$Yb nucleus, while $\sigma$  is the quality measure Eq \ref{diseqn73} . } 
\label{tab2}
\end{longtable}
% 163 Dy %
%\begin{longtable}[H]
%\centering

\begin{longtable}{ccccccc}
%\toprule
\hline \hline \hline
\qquad \qquad \qquad  $ B_{\beta}\neq B_{\gamma}\neq B_{rot}  $&& & & & $ B_{\beta}= B_{\gamma}= B_{rot} $\\
\hline
  &  & $a$=0\quad \quad   & DDM \quad \quad \quad \quad \quad \quad \quad &$ a$=0& DDM &  \\ \hline \hline
%\midrule
 g.s \\
$\sigma$ &  & 0.0191 & 0.0191\quad \quad \quad \quad \quad \quad \quad & 1.4517& 0.0362 & \\ 
a &  & \quad \quad & 0.0002\quad \quad \quad \quad \quad \quad \quad & & 0.0041 & \\
$\beta_{0}$ &  & \quad \quad & 0.0527\quad \quad \quad \quad \quad \quad \quad & & 5.5363 & \\\hline \hline
 $\beta$ \\  
$\sigma$ &  & 0.2462 & 0.2462\quad \quad \quad \quad \quad \quad \quad &1.0349 & 0.2283 & \\ 

a &  & \quad \quad & 0.00011\quad \quad \quad \quad \quad \quad \quad & & 0.004 & \\
$\beta_{0}$ &  & \quad \quad & 0.6614\quad \quad \quad \quad \quad \quad \quad & & 2.082 & \\\hline

\hline \hline
$\sigma_{total}$ &  & 0.1113 & 0.1113\quad \quad \quad \quad \quad \quad \quad & 1.2997& 0.7706 & \\ 
a &  & \quad \quad & 0.00005\quad \quad \quad \quad \quad \quad \quad & & 0.0072 & \\
$\beta_{0}$ &  & \quad \quad & 1.0456\quad \quad \quad \quad \quad \quad \quad & & 2.2852 &\\\hline

\hline \hline
%\bottomrule
%\end{longtable}
\caption {   $\beta_{0}$ and a the position of the minimum of Davidson potential  and the deformation dependence of the mass parameter  respectively for the $^{163}$Dy nucleus, while $\sigma$  is the quality measure Eq \ref{diseqn73} .}
\label{tab3}
\end{longtable}
%\begin{longtable}[H]
%\centering
\newpage
\begin{longtable}{ccccccc}
%\toprule
\hline \hline \hline
\qquad \qquad \qquad   $B_{\beta}\neq B_{\gamma}\neq B_{rot}$  && & & &  $B_{\beta}= B_{\gamma}= B_{rot}$ \\
\hline
L  & Exp. \cite{22} & $a$=0\quad \quad   & DDM \quad \quad \quad \quad \quad \quad \quad & $a$=0& DDM &  \\ \hline \hline
%%\midrule
 g.s & & \quad \quad  & \quad \quad \quad \quad \quad \quad \quad &&  &  \\
$9/2^-$ & 2.27 & 2.28\quad \quad  & 2.28\quad \quad \quad \quad \quad \quad \quad &2.25  &2.28  &\\
$11/2^-$ & 3.84 & 3.84\quad \quad  & 3.84\quad \quad \quad \quad \quad \quad \quad &3.74 &3.83  &  \\
$13/2^-$ & 5.68 & 5.67\quad \quad  & 5.67\quad \quad \quad \quad \quad \quad \quad &5.43 & 5.66 &  \\
$15/2^-$ & 7.78 & 7.76\quad \quad  &7.77\quad \quad \quad \quad \quad \quad \quad &7.32 &7.76  &  \\
$17/2^-$ & 10.07 & 10.12\quad \quad  & 10.13\quad \quad \quad \quad \quad \quad \quad &9.37 & 10.12 &  \\
$19/2^-$ &12.76 & 12.73\quad \quad  & 12.75\quad \quad \quad \quad \quad \quad \quad &11.58 &12.74  &  \\
$21/2^-$ & 15.63 & 15.58\quad \quad  & 15.63\quad \quad \quad \quad \quad \quad \quad &13.91 &15.62  &  \\
$23/2^-$ & 18.75 & 18.68\quad \quad  & 18.74\quad \quad \quad \quad \quad \quad \quad &16.37 &18.76  &  \\
%25/2^- & & 22.01\quad \quad & 22.10\quad \quad \quad \quad \quad \quad \quad & 18.92& 22.16 & 21.63 \\
%27/2^- &25.71 &25.56\quad \quad & 25.68\quad \quad \quad \quad \quad \quad \quad & 21.56& 25.81 & 28.52 
 
 \\\hline \hline
 $\beta$ & & \quad \quad  & \quad \quad \quad \quad \quad \quad \quad &  \\
$9/2^-$ & 13.46 & 13.44\quad \quad  & 13.42\quad \quad \quad \quad \quad \quad \quad & 14.40 & 14.40 &  \\
$11/2^-$ & 14.75 & 15.00\quad \quad  & 14.98\quad \quad \quad \quad \quad \quad \quad & 15.89& 15.89 &  \\
$13/2^-$ & 16.38 & 16.83\quad \quad  & 16.81\quad \quad \quad \quad \quad \quad \quad & 17.58& 17.58 &  \\
$15/2^-$ & 19.48 &18.92\quad \quad  &18.90\quad \quad \quad \quad \quad \quad \quad & 19.47&19.47 &  \\
$17/2^-$ & 20.73 & 21.28\quad \quad  & 21.26\quad \quad \quad \quad \quad \quad \quad & 21.52& 21.52 &  \\
$19/2^-$ &23.27 & 23.89\quad \quad  & 23.87\quad \quad \quad \quad \quad \quad \quad & 23.73& 23.73 &  \\
$21/2^-$ & 27.99 &26.74\quad \quad  & 26.73\quad \quad \quad \quad \quad \quad \quad & 26.06& 26.06 &  \\
%$23/2^-$ & 30.63 & 29.84\quad \quad  & 29.83\quad \quad \quad \quad \quad \quad \quad & 28.52& 28.52 &  \\
%25/2^- &33.15 & 33.17\quad \quad & 33.16\quad \quad \quad \quad \quad \quad \quad & 31.07& 31.07 &  

%27/2 &0 &1\quad \quad & 5\quad \quad \quad \quad \quad \quad \quad & 0.00& 2 & 6 \\ \hline
 \\\hline \hline
 $\gamma$ & & \quad \quad  & \quad \quad \quad \quad \quad \quad \quad &  \\
$9/2^-$ & 18.59 & 18.69\quad \quad  & 18.69\quad \quad \quad \quad \quad \quad \quad &19.94  &20.09  &  \\
$11/2^-$ & 20.61 & 20.16\quad \quad  & 20.16\quad \quad \quad \quad \quad \quad \quad &21.01 &21.18  &  \\
$13/2^-$ & 22.22 & 21.88\quad \quad  & 21.88\quad \quad \quad \quad \quad \quad \quad &22.26 &22.45  & \\
$15/2^-$ & 23.40 & 23.86\quad \quad  &23.86\quad \quad \quad \quad \quad \quad \quad &23.67 &23.88  &  \\
$17/2^-$ & 25.85 & 26.09\quad \quad  & 26.09\quad \quad \quad \quad \quad \quad \quad &25.23 &25.47  &  \\
$19/2^-$ &28.56 & 28.56\quad \quad  & 28.56\quad \quad \quad \quad \quad \quad \quad &26.93 &27.21  & \\
%$21/2^-$ & 30.44 & 31.27\quad \quad  & 31.27\quad \quad \quad \quad \quad \quad \quad & 28.77& 29.09 &  \\
%23/2^{-} & 32.31 & 34.21\quad \quad  & 34.21\quad \quad \quad \quad \quad \quad \quad & 30.72& 31.09 & 

%25/2 &0 & 1\quad \quad & 2\quad \quad \quad \quad \quad \quad \quad & 0.00& 2 & 6 \\
%27/2 &0 &1\quad \quad & 5\quad \quad \quad \quad \quad \quad \quad & 0.00& 2 & 6 \\ \hline

 \\ \hline
\hline \hline
%\bottomrule
%\end{longtable}
\caption { Comparison of the theoretical predictions of energy levels \ref{diseqn46} of the ground state band,the $\beta$ and $\gamma$ bands normalized to the energy of the first excited state $E(7/2^-_{\text{g}.s.})$ using the parameters given in Tables \ref{tab1} and \ref{tab2} for $^{173}Yb$ for this work with experimental values taken from Ref \cite{22}.}
\label{tab5}
\end{longtable}

\newpage

% 163 Dy %
%\begin{longtable}[H]
%\centering
\begin{longtable}{cccccc}
%\toprule
\hline \hline \hline
\qquad \qquad \qquad \quad  $B_{\beta}\neq B_{\gamma}\neq B_{rot}$    & &&  \quad \quad \quad \quad    $ B_{\beta}= B_{\gamma}= B_{rot} $\\
\hline
L& Exp. \cite{22} &$ a=0$  \quad \quad \quad \quad   \quad \quad & $a=0$&DDM &  \\ \hline \hline
%%\midrule
 g.s &  \quad \quad  &   \quad \quad  &&  &  \\
$9/2^-$ & 2.28\quad \quad & 2.28\quad \quad \quad \quad   \quad \quad  & 2.24 & 2.28  &\\
$11/2^-$ & 3.84\quad \quad & 3.83\quad \quad \quad \quad   \quad \quad  &3.71  &3.83  &\\
$13/2^-$ & 5.66\quad \quad & 5.66\quad \quad \quad \quad   \quad \quad  &5.37  &5.64  &\\
$15/2^-$ & 7.75\quad \quad & 7.74\quad \quad \quad \quad   \quad \quad    &7.20  &7.72  &\\
$17/2^-$ & 10.13 \quad \quad& 10.08\quad \quad \quad \quad   \quad \quad    &9.18  &10.06  &\\
$19/2^-$ &12.68\quad \quad & 12.67\quad \quad \quad \quad   \quad \quad     &11.28  &12.65  &\\
$21/2^-$ & 15.49\quad \quad & 15.51\quad \quad \quad \quad   \quad \quad    &13.50  &15.50  &\\
$23/2^-$ & 18.56\quad \quad & 18.57\quad \quad \quad \quad   \quad \quad    &15.82  &18.60  &\\
%25/2^- &21.82 & 21.86\quad \quad & 21.86\quad \quad \quad \quad \quad \quad \quad & 18.22&21.95 & 19.17 \\
%27/2^- &25.36 &25.36\quad \quad & 25.36\quad \quad \quad \quad \quad \quad \quad & 20.69& 25.55 & 21.88

 \\\hline \hline
 $\beta$ & & \quad \quad  & \quad \quad \quad \quad \quad \quad \quad &  \\
$5/2^-$ & 9.70 & 9.93\quad \quad \quad \quad   \quad \quad    &10.76  &9.90  &\\
$7/2^- $& 10.91 & 10.93\quad \quad \quad \quad   \quad \quad   &11.76  &10.93  &\\
$9/2^-$ & 12.47 & 12.21\quad \quad \quad \quad   \quad \quad   &13.01  &12.22  &\\
$11/2^-$ &  & 13.77\quad \quad \quad \quad   \quad \quad   &14.47 &13.75  &\\
$13/2^-$ &  & 15.59\quad \quad \quad \quad   \quad \quad    &16.13  &15.49  &\\
$15/2^-$ & & 17.68\quad \quad \quad \quad   \quad \quad    &17.96  &17.44  &\\
$17/2^-$ &  &20.02\quad \quad \quad \quad   \quad \quad    &19.94  &19.56  &\\
$19/2^-$ &  & 22.61\quad \quad \quad \quad   \quad \quad    &22.04  &21.85  &\\
$21/2^-$ &  & 25.44\quad \quad \quad \quad   \quad \quad     &24.26  &24.28 &\\
$23/2^-$ &  & 28.50\quad \quad \quad \quad   \quad \quad    &26.58  &26.85  &\\
$25/2^-$ & & 31.79\quad \quad \quad \quad   \quad \quad  &28.98  &29.55  &
%27/2 &0 &1\quad \quad & 5\quad \quad \quad \quad \quad \quad \quad & 0.00& 2 & 6 \\ \hline

 \\\hline 
\hline \hline
%\bottomrule
%\end{longtable}
\caption {Comparison of the theoretical predictions of energy levels \ref{diseqn46} of the ground state and $\beta$  bands normalized to the energy of the first excited state $E(7/2^-_{\text{g}.s.})$ using the parameters given in Tables \ref{tab1} and \ref{tab3} for $^{163}Dy$ for this work with experimental values taken from Ref \cite{22} .}
\label{tab6}
\end{longtable}
 
\newpage

%\begin{longtable}[H]
%\centering
\begin{longtable}{ccccccc}
%\toprule
\hline \hline \hline
\qquad \qquad \qquad  $ B_{\beta}\neq B_{\gamma}\neq B_{rot} $ && & & &  $B_{\beta}= B_{\gamma}= B_{rot}$ \\
\hline
& Exp. \cite{22} & $a=0$\quad \quad   & DDM \quad \quad \quad \quad \quad \quad \quad & $a=0$& DDM &  \\ \hline \hline
%\midrule \\
 $\frac{B(E2;L'_{\text{g.s.}} \rightarrow L_{\text{g.s.}})}{B(E2;\frac{9}{2}^-_{\text{g.s.}} \rightarrow \frac{5}{2}^-_{\text{g.s.}})}$ & & \quad \quad  & \quad \quad \quad \quad \quad \quad \quad &&  &  \\
$11/2^-\longrightarrow7/2^- $& 2.03 & 1.70\quad \quad  & \quad \quad \quad \quad \quad \quad \quad &1.74  &1.72  & \\
$13/2^-\longrightarrow9/2^-  $& 2.06 & 2.18\quad \quad  & \quad \quad \quad \quad \quad \quad \quad &2.27 &2.23  &  \\
$15/2^-\longrightarrow11/2^- $ & 2.31 & 2.51\quad \quad  & \quad \quad \quad \quad \quad \quad \quad &2.68 & 2.60 & \\
$17/2^-\longrightarrow13/2^- $ & 2.93 & 2.76\quad \quad  &\quad \quad \quad \quad \quad \quad \quad &3.02 &2.89  &  \\
$21/2^-\longrightarrow17/2^- $ & 3.21 & 3.10\quad \quad  & \quad \quad \quad \quad \quad \quad \quad &3.60 &3.33  & \\
$23/2^-\longrightarrow19/2^- $ &3.26 & 3.23\quad \quad  & \quad \quad \quad \quad \quad \quad \quad &3.86 &3.51  &  \\
$25/2^-\longrightarrow21/2^-$ & 3.37 & 3.34\quad \quad  & \quad \quad \quad \quad \quad \quad \quad &4.11 &3.66 &  
\\\hline \hline \\ 
 $\frac{B(E2;L'_{\beta} \rightarrow L_{\text{g.s.}})}{B(E2;\frac{9}{2}^-_{\text{g.s.}} \rightarrow \frac{5}{2}^-_{\text{g.s.}})}\times 10^3$ & & \quad \quad  & \quad \quad \quad \quad \quad \quad \quad &  \\
$9/2^-\longrightarrow 5/2^-$  &  & 1.74\quad \quad  & \quad \quad \quad \quad \quad \quad \quad & 11.81 & 11.81 &  \\
$13/2^-\longrightarrow9/2^- $ &  & 1.66\quad \quad  & \quad \quad \quad \quad \quad \quad \quad & 14.47& 14.47 &  \\
$17/2^-\longrightarrow13/2^- $ &  & 0.55\quad \quad  &\quad \quad \quad \quad \quad \quad \quad & 9.78& 9.78 &  \\
$9/2^-\longrightarrow9/2^- $ &  & 0.76\quad \quad  &\quad \quad \quad \quad \quad \quad \quad & 4.65& 4.65 &  \\
$13/2^-\longrightarrow13/2^- $ &  & 4.82\quad \quad  & \quad \quad \quad \quad \quad \quad \quad & 29.55& 29.55 &  \\
$17/2^-\longrightarrow17/2^- $ & & 7.45\quad \quad  & \quad \quad \quad \quad \quad \quad \quad & 45.80& 45.81 &  \\
$5/2^-\longrightarrow9/2^-  $&  & 16.63\quad \quad  & 16.61\quad \quad \quad \quad \quad \quad \quad & 97.81& 97.81 &  \\
$9/2^-\longrightarrow13/2^- $ &  & 39.24\quad \quad  & 39.21\quad \quad \quad \quad \quad \quad \quad & 220.25& 220.25 &  \\
$13/2^-\longrightarrow17/2^-  $& & 56.40\quad \quad & 56.35\quad \quad \quad \quad \quad \quad \quad & 296.76& 296.77 &  
%27/2 &0 &1\quad \quad & 5\quad \quad \quad \quad \quad \quad \quad & 0.00& 2 & 6 \\ \hline
\\\hline \hline \\
 $\frac{B(E2;L'_{\gamma} \rightarrow L_{\text{g.s.}})}{B(E2;\frac{9}{2}^-_{\text{g.s.}} \rightarrow \frac{5}{2}^-_{\text{g.s.}})}\times 10^3$ & & \quad \quad  & \quad \quad \quad \quad \quad \quad \quad &  \\
$9/2^-\longrightarrow5/2^-  $&  & 9.19\quad \quad  & \quad \quad \quad \quad \quad \quad \quad & 41.03 & 41.30 &  \\
$9/2^-\longrightarrow9/2^-$  &  & 1.42\quad \quad  & \quad \quad \quad \quad \quad \quad \quad &6.79& 6.82 &  \\
$9/2^-\longrightarrow13/2^-  $&  & 28.84\quad \quad  & \quad \quad \quad \quad \quad \quad \quad & 150.41& 150.92 & \\
$11/2^-\longrightarrow9/2^- $ &  & 21.66\quad \quad  &\quad \quad \quad \quad \quad \quad \quad & 102.63& 103.23 &  \\
$11/2^-\longrightarrow13/2^- $ &  & 19.95\quad \quad  & \quad \quad \quad \quad \quad \quad \quad & 103.17& 103.57 &  \\
$13/2^-\longrightarrow9/2^- $ & & 19.91\quad \quad  & \quad \quad \quad \quad \quad \quad \quad & 93.00& 93.61 & \\
$13/2^-\longrightarrow13/2^-$  &  & 9.10\quad \quad  & \quad \quad \quad \quad \quad \quad \quad & 46.53& 46.75 &  \\
$13/2^-\longrightarrow17/2^-$  &  & 33.70\quad \quad  & \quad \quad \quad \quad \quad \quad \quad & 188.55& 189.04 &  \\
$15/2^-\longrightarrow13/2^-  $&  & 13.29\quad \quad  & \quad \quad \quad \quad \quad \quad \quad & 67.06& 67.42 & 
%25/2 &0 & 1\quad \quad & 2\quad \quad \quad \quad \quad \quad \quad & 0.00& 2 & 6 \\
%27/2 &0 &1\quad \quad & 5\quad \quad \quad \quad \quad \quad \quad & 0.00& 2 & 6 \\ \hline
\\\hline \hline \hline
%\bottomrule
%\end{longtable}
\\
\caption{Comparison of the theoretical predictions of B(E2) in units of $B(E2;9/2^-_{\text{g.s.}} \rightarrow 5/2^-_{\text{g.s.}})$ using the parameters
	given in Tables \ref{tab1} and  \ref{tab2}  for $^{173}Yb$ in this work with experimental values \cite{22}.
}
\label{tab8}
\end{longtable}
%\begin{longtable}[H]
%\centering
\newpage
\begin{longtable}{ccccccc}
%\toprule
\hline \hline \hline
\qquad \qquad \qquad  $ B_{\beta}\neq B_{\gamma}\neq B_{rot}  $&& & & & $ B_{\beta}= B_{\gamma}= B_{rot} $\\
\hline
&Exp. \cite{22}   & $a=0$& DDM \quad \quad   \quad & a=0&DDM & \\ \hline \hline 
%\midrule 
 $\frac{B(E2;L'_{\text{g.s.}} \rightarrow L_{\text{g.s.}})}{B(E2;\frac{9}{2}^-_{\text{g.s.}} \rightarrow \frac{5}{2}^-_{\text{g.s.}})}$ & & \quad \quad  & \quad \quad \quad \quad \quad \quad \quad &&  &  \\ 
$11/2^-\longrightarrow7/2^- $ & 1.63 & 1.71  & \quad \quad \quad  &1.75  &1.73  & \\
$13/2^-\longrightarrow9/2^-  $& 2.04 & 2.18  & \quad \quad \quad  &2.30 & 2.25 &  \\
$15/2^-\longrightarrow11/2^- $ & 2.80 & 2.52  & \quad \quad \quad  &2.73 &2.63  &  \\
$17/2^-\longrightarrow13/2^- $ & 2.44 & 2.77  &\quad \quad \quad  &3.10 &2.94  & \\
$19/2^-\longrightarrow15/2^-  $& 2.6 & 2.96  &\quad \quad \quad  &3.43 &3.20  &  \\
$21/2^-\longrightarrow17/2^- $ & 2.44 & 3.12 & \quad \quad \quad  & 3.73& 3.41 &  \\
$23/2^-\longrightarrow19/2^-  $ & 2.28 & 3.26  & \quad \quad \quad  &4.03 &3.61  &  \\

\\\hline \hline \\
 $\frac{B(E2;L'_{\beta} \rightarrow L_{\text{g.s.}})}{B(E2;\frac{9}{2}^-_{\text{g.s.}} \rightarrow \frac{5}{2}^-_{\text{g.s.}})}\times 10^3$ & & \quad \quad  & \quad \quad \quad \quad \quad \quad \quad &  \\
$9/2^-\longrightarrow 5/2^- $ & & 1.55  & \quad \quad \quad  & 11.17 & 15.02 &   \\
$13/2^-\longrightarrow9/2^- $ & & 1.17  & \quad \quad \quad  & 12.36& 19.70 &  \\
$9/2^-\longrightarrow9/2^-   $& & 0.81 &\quad \quad \quad  & 5.16& 5.85 &  \\
$13/2^-\longrightarrow13/2^-$  & & 5.13  & 5.12\quad \quad \quad  & 32.83& 37.49 & \\
$17/2^-\longrightarrow17/2^-$ & & 7.93 & \quad \quad \quad  & 50.91& 58.62 &  \\
$9/2^-\longrightarrow13/2^-  $ & & 45.67  & 45.66\quad \quad \quad &264.08 & 271.78 &   \\
$13/2^-\longrightarrow17/2^-$ & & 66.47 & 66.45\quad \quad \quad  &354.9& 360.08 &  \\
%27/2 &0 &1\quad \quad & 5\quad \quad \quad \quad \quad \quad \quad & 0.00& 2 & 6 \\ \hline
\hline \hline \hline
%\bottomrule
%\end{longtable}
\\
\caption{Comparison of the theoretical predictions of B(E2) in units of $B(E2;9/2^-_{\text{g.s.}} \rightarrow 5/2^-_{\text{g.s.}})$ using the parameters
	given in Tables \ref{tab1} and  \ref{tab3} for $^{163}Dy$ in this work with experimental values \cite{22}.
}
\label{tab9}
\end{longtable}

\newpage
\appendix{}
\section{ Special cases of energy spectrum \label{appendixA} }
\subsection*{ Special case 1: Without mass coefficients}
If we assume  $B_{\beta}$=$B _{\gamma}$=$B _{rot}$ and $\hbar= 1$, we get from
 Eq. \eqref{diseqn42}
 \begin{align}
 &k_2=a^2\Big[5(1-\delta-\lambda)+(1-2\delta)(1-2\lambda)+6+\Lambda\Big]+\frac{2V_{0}}{\beta^2_{0}},\nonumber
 \\
 &k_0=a\Big[5(1-\delta-\lambda)+8+2\Lambda\Big]-4V_{0},\label{diseqn48}\\
 &k_{-2}=2+\Lambda+2V_{0}\beta^2_{0}. \nonumber
 \end{align}
 Consequently, the energy spectrum formula Eq. \ref{diseqn46} is identical
 to Eq. (82) of Ref. \cite{5} obtained by means of supersymmetric
 quantum mechanical method (SUSYQM) \cite{17,18}. The slight difference between our coefficients $k_{2}$, $k_{0}$ and
$ k_{-2}$ and those of Ref. \cite{5} comes from the adopted expression  of Davidson potential.
 \subsection*{Special case 2: No dependence of the mass on
 	the deformation}
 If $a = 0$, i.e., the dependence of the mass on the deformation
 is canceled, then one has from Eq. \eqref{diseqn42}
 \begin{align}
& k_2=\frac{2g_{\beta}}{\beta^4_{0}}, \quad \quad k_0=-4\frac{g_{\beta}}{\beta^2_{0}},   \nonumber\\
&k_{-2}=\frac{B_{\beta}}{B_{\gamma}}\Big(1+\frac{B_{\beta}}{B_{\gamma}}\Big)+\frac{B_{\beta}}{\hbar^2}\Lambda+2g_{\beta}.
 \label{diseqn49}
 \end{align}
 In this case, the energy spectrum formula reads
 \begin{align}
 E_{n_{\beta}n_{\gamma}L|m|}=\sqrt{2\frac{V^2_{0}}{g_{\beta}}}\Bigg[1+2n_{\beta}+\frac{1}{2}q_{n_{\gamma}}(L,|m|)-\sqrt{2g_{\beta}}\Bigg],
 \label{diseqn51}
 \end{align} 
 with
  \begin{align}
 \frac{1}{2}q_{n_{\gamma}}(L,|m|)=\sqrt{\frac{1}{4}+\frac{B_{\beta}}{B_{\gamma}}\Bigg(1+\frac{B_{\beta}}{B_{\gamma}}\Bigg)+\frac{B_{\beta}}{\hbar^2}\Lambda+2g_{\beta}},
  \label{diseqn52}
  \end{align}
 and
 \begin{align}
 \frac{B_{\beta}}{\hbar^2}\Lambda&=\frac{2}{g}\frac{B_{\beta}}{B_{\gamma}}(1+n_{\gamma})+\frac{m^2}{3}\frac{B_{\beta}}{B_{\gamma}}+\frac{1}{3}\frac{B_{\beta}}{B_{rot}}\Big(L(L+1)\nonumber\\+&j(j+1)-K^2-(K-2m)^2\Big)\label{diseqn53}\\-&\frac{1}{3\xi}\Big(3(K-2m)^2-j(j+1)\Big),\nonumber
 \end{align}
 note that Eq. \eqref{diseqn51} represents the correct formula of the
 energy spectrum, compared to Eq. (11) given in Ref. \cite{9},
 where the mass parameter term is missed in the analog formula of Eq. \eqref{diseqn52}.
  \subsection*{ Special case 3: Standard case}  
 In the case of $a = 0$ and  $B_{\beta}$=$B _{\gamma}$=$B _{rot}$, our formula Eq. \eqref{diseqn46} becomes
 \begin{equation}
 E_{n_{\beta}n_{\gamma}L|m|}=\sqrt{2\frac{V^2_{0}}{g_{\beta}}}\Big(1+2n_{\beta}+\sqrt{\frac{9}{4}+\Lambda+2g_{\beta}}\Big)-2V_{0},
 \label{diseqn54}
 \end{equation} 
the Eq. \eqref{diseqn54} represents the correct formula of the energy spectrum, compared to Eq. (13) given in Ref. \cite{10}.
\section{Formulas used for the calculations of the B(E2) transitions \label{appendixb}}
In this Appendix we present the expressions used for calculations of the transitions probabilities B(E2) :
\begin{align}
&\frac{B(E2;L'_{\text{g.s.}} \rightarrow L_{\text{g.s.}})}{B(E2;\frac{9}{2}_{\text{g.s.}} \rightarrow \frac{5}{2}_{\text{g.s.}})}=\Bigg(\frac{(L'2K0|LK)}{(\frac{9}{2}2\frac{5}{2}0|\frac{5}{2}\frac{5}{2})}\Bigg)^2  \nonumber\\ &\times\Bigg(\frac{\Gamma[0.25(q_{0}(L',0)+q_{0}(L,0))+1.5]}{\Gamma[0.25(q_{0}(\frac{9}{2},0)+q_{0}(\frac{5}{2},0))+1.5]}\Bigg)^2 \label{B1}\\ &\times\Bigg(\frac{\Gamma[0.5q_{0}(\frac{9}{2},0)+1]\Gamma[0.5q_{0}(\frac{5}{2},0)+1]}{\Gamma[0.5q_{0}(L',0)+1]\Gamma[0.5q_{0}(L,0)+1]}\Bigg), \nonumber
\end{align}
\begin{align}
&\frac{B(E2;L'_{\beta} \rightarrow L_{\text{g.s.}})}{B(E2;\frac{9}{2}_{\text{g.s.}} \rightarrow \frac{5}{2}_{\text{g.s.}})}=\frac{1}{4}\times\Bigg(\frac{(L'2K0|LK)}{(\frac{9}{2}2\frac{5}{2}0|\frac{5}{2}\frac{5}{2})}\Bigg)^2\nonumber\\ &\times\Bigg(\frac{\Gamma[0.25(q_{0}(L',0)+q_{0}(L,0))+1.5]}{\Gamma[0.25(q_{0}(\frac{9}{2},0)+q_{0}(\frac{5}{2},0))+1.5]}\Bigg)^2 \label{B2}\\ &\times\Bigg(\frac{\Gamma[0.5q_{0}(\frac{9}{2},0)+1]\Gamma[0.5q_{0}(\frac{5}{2},0)+1]}{\Gamma[0.5q_{0}(L',0)+1]\Gamma[0.5q_{0}(L,0)+1]}\Bigg)\nonumber\\  &\times \frac{(0.5q_{0}(L',0)-0.5q_{0}(L,0)-1)^2}{0.5q_{0}(L',0)+1},
\nonumber
\end{align}
\begin{align}
\frac{B(E2;L'_{\gamma} \rightarrow L_{\text{g.s.}})}{B(E2;\frac{9}{2}_{\text{g.s.}} \rightarrow \frac{5}{2}_{\text{g.s.}})}=g\times\Bigg(\frac{(L'2K0|LK)}{(\frac{9}{2}2\frac{5}{2}0|\frac{5}{2}\frac{5}{2})}\Bigg)^2 \nonumber\\ \times\Bigg(\frac{\Gamma[0.25(q_{1}(L',1)+q_{0}(L,0))+1.5]}{\Gamma[0.25(q_{0}(\frac{9}{2},0)+q_{0}(\frac{5}{2},0))+1.5]}\Bigg)^2 \label{B3}\\ \times\Bigg(\frac{\Gamma[0.5q_{0}(\frac{9}{2},0)+1]\Gamma[0.5q_{0}(\frac{5}{2},0)+1]}{\Gamma[0.5q_{1}(L',1)+1]\Gamma[0.5q_{0}(L,0)+1]}\Bigg),
\nonumber
\end{align}
where $(L'2K0|LK)$ is Clebsch-Gordan coefficient.
% BibTeX users please use
% \bibliographystyle{}
% \bibliography{}

\end{document}